\documentclass{aa}
\usepackage{txfonts}
\usepackage{graphicx}
\usepackage{epsfig}
\usepackage{natbib}
\bibpunct{(}{)}{;}{a}{}{,}

\def\grsim{\mathrel{\hbox{\rlap{\hbox{\lower4pt\hbox{$\sim$}}}\hbox{$>$}}}}

\begin{document}

\title{On emission-line spectra obtained from evolutionary synthesis
models} \subtitle{I. Dispersion in the ionising flux and Lowest Luminosity Limits}
\author{M. Cervi\~no\inst{1} \and V. Luridiana\inst{1} \and
E. P\'erez\inst{1} \and J.M. V\'{\i}lchez\inst{1} \and D. Valls-Gabaud\inst{2}}

\institute{Instituto de Astrof\'{\i}sica de Andaluc{\'{\i}}a (CSIC), 
           Camino bajo de Hu\'etor 24, Apdo. 3004, 18080 Granada, Spain
  \and     UMR CNRS 5572, Laboratoire d'Astrophysique,
           Observatoire Midi-Pyr\'en\'ees, 14 Avenue Edouard Belin, 
           31400 Toulouse, France
}

\offprints{Miguel Cervi\~no}
\mail{mcs@laeff.esa.es}
\date{Received April 1, 2003; accepted June 11, 2003}
\authorrunning{Cervi\~no et al.}
\titlerunning{Emission-line spectra and synthesis models}

\abstract{
Stellar clusters with the same 
general physical properties (e.g., total mass, age, 
and star-formation mode) may have very different 
stellar mass spectra due to the incomplete sampling
of the underlying mass function;
such differences are especially relevant in the
high-mass tail of the mass function due to the 
smaller absolute number
of massive stars.
Since the ionising spectra of star-forming regions are 
mainly produced by massive stars and their by-products,
the dispersion in the number of massive stars across individual 
clusters also produces a dispersion in the properties of the 
corresponding ionising spectra.
This implies that regions with the same physical properties may
produce very different emission line spectra, 
and occupy different positions in emission-line diagnostic diagrams.  
In this paper, we lay the bases for the future analysis of this effect 
by evaluating the dispersion in the
ionising fluxes of synthetic spectra computed with evolutionary models.
As an important consequence of the explicit consideration of
sampling effects, we found that the intensities of synthetic fluxes 
at different ionisation edges are strongly correlated,
a fact suggesting that no additional dispersion will result from
the inclusion of sampling effects in
the analysis of diagnostic diagrams;
this is true for \ion{H}{ii} regions on all scales,
those ionised by single massive stars through those
ionised by super stellar clusters.
This finding is especially relevant, in consideration
of the fact that real \ion{H}{ii} regions are found in a 
band sensibly narrower than predicted by standard methods.
Additionally, we find convincing suggestions that 
the \ion{He}{ii} line intensities 
are strongly affected by sampling, 
especially during the WR phase, 
and so cannot be used to constrain the evolutionary status of
stellar clusters.  
We also establish the range of applicability of synthesis models
set by the Lowest Luminosity Limit for the ionising flux, 
that is the lowest limit in cluster mass 
for which synthesis models can be applied to predict ionising spectra.
This limit marks the boundary between the situations in which the ionising
flux is better modeled with a single star as opposed to 
a star cluster;
this boundary depends on the metallicity and age of the stellar
population, ranging from 10$^3$ to more than 10$^6$ M$_\odot$ 
in the case of a single burst event.
As a consequence, synthesis models should not be used 
to try to account for the properties of clusters with smaller masses.
 } \maketitle

\section{Introduction and motivation}\label{sec:introduction}

In recent years a great deal of effort has been put into modeling more
realistic ionising fluxes of massive stars \cite[e.g., the {\sc wm-basic} code
by][and references therein]{WMbasic}.  These
new model atmospheres, as included in evolutionary synthesis codes, are
expected to produce more reliable ionising spectra, which in turn should
allow a better interpretation of the emission line spectra of ionised
nebulae, through the use of photoionisation codes \cite[e.g.,][]{linda}.

Yet the ionising spectra of stellar clusters are produced by the most
massive stars, whose absolute numbers can be strongly affected by 
the incomplete sampling of the underlying Initial Mass Function (IMF). 
The effects of incomplete sampling on the observed properties
of star-forming regions can be currently evaluated in two main ways:
either by means of Monte Carlo synthesis models,
or by means of an appropriate statistical formalism 
applied to the results of analytical synthesis models.
By analytical synthesis models, or analytical models
for short, we will in the following refer
to those population synthesis models computed 
by codes that neglect the issue of the incomplete sampling, 
and assume instead that the mass spectrum of stellar clusters 
can be represented by a continuous, analytical function that coincides
with the underlying IMF. 
On the other hand, Monte Carlo codes simulate stellar populations
by the stochastic generation of stars, which is stopped when either
a given number of stars or a given cluster mass is reached; 
the mass probability function for each star is given by the IMF,
in such a way that the actual mass spectrum approaches the IMF
as the number of stars increases,
whereas large deviations occur for small clusters.
This is assumed to be a more realistic representation of real
stellar clusters than the one provided by analytical codes.

The objective of this work is to study quantitatively 
the impact on emission line spectra
of the sampling effects in the IMF,
and to evaluate its consequences in the determination of
the physical properties of \ion{H}{ii} regions and  galaxies.
This is not a trivial problem. A simple, perhaps na\"{\i}ve, approach would 
go through the following steps:
{\sl (i)} building synthetic clusters with different numbers of stars with a 
stochastic
sampling of the IMF, ensuring in this way that only integer numbers of stars are
included; {\sl (ii)} obtaining the corresponding Spectral Energy Distributions
(SEDs); {\sl (iii)} using the SEDs as inputs to a photoionisation code;  and {\sl (iv)} 
drawing conclusions based on the overall results.
However, since the number of free parameters is pretty large, it is 
far more useful to split the problem into several smaller steps, 
each of which addressing a particular source of uncertainty. A possible 
procedure adopting this strategy is the following:

\begin{itemize}

\item A first step is to evaluate quantitatively the sampling effects
on the ionising spectrum. Although the computation of photoionisation models
is not involved in this step, it is important to characterize the resulting ionising flux 
with a view on the physics of photoionisation.

\item A second step is to determine how the different emission lines
scale with the size (mass/number of stars) of the stellar cluster.  
For example, the
intensity of hydrogen recombination lines roughly scale 
with the number of ionising
photons above the threshold energy, and hence with the number of massive
stars in the cluster (for a given cluster age and given properties of the
emitting gas).  However, a similar scaling relation  -provided it exists- has not
yet been established for the case of forbidden emission lines.

\item A third step is to investigate the influence of both the stellar
and the nebular properties on the results of
photoionisation models.
A large grid of photoionisation models would
be required to explore exhaustively the parameter space, both along stellar
population dimensions (i.e., IMF slope and mass limits, ages, stellar
metallicity, model atmospheres, etc.), and along  gas properties dimensions
(density profile, chemical abundances, radius of the nebulae, covering
factor, etc.). In this step we can partially rely on extensive studies 
performed by different groups \cite[e.g.][ and references therein]{Dop00}.
Feedback from observations must also be taken into account.

\end{itemize}

In this series of papers, we address each of these problems in turn.  This
first paper presents a quantitative evaluation of sampling effects on the
predicted ionising flux for a wide range of input parameters in synthesis models,
allowing us to obtain some conditions necessary, but not sufficient, 
to explain the correlations of observational data in emission-line 
diagnostic diagrams.
In addition, we determine a mass limit under which 
stellar clusters should be better modeled with single
individual stars rather than with stellar clusters.
In a second paper we will determine the scaling relations of the
emission-line spectrum with the initial mass/number of stars in the
cluster.  The third, and last, paper of this series will be devoted to the
analysis of a set of Monte Carlo simulations of evolutionary synthesis
models linked with a photoionisation code, where some specific properties
will be assumed for the emitting gas. Throughout all these papers we will
discuss the relationships between sampling effects and the ionising continuum, along
with the resulting emission-line spectra.

The importance of observational feedback in this analysis cannot be
overstated.  
A long-standing puzzle is posed by 
the positions in emission-line diagnostic
diagrams of observed \ion{H}{ii} regions
ionised both by single stars and by stellar clusters 
cover an area narrower than the one expected on the basis
of extensive grids of
photoionisation models.
For example, \cite{Dop00} built a grid of zero-age stellar clusters 
varying the metallicity and the ionisation parameter of the model nebulae,
and found that the area covered on diagnostic diagrams 
by these models is much broader than the area covered by observational data, 
suggesting the existence of a hidden
parameter that constrains the possible positions of real \ion{H}{ii} regions
in diagnostic diagrams.  
This problem has also been thoroughly discussed by \cite{SI03};
these authors, who also consider the evolution of the
ionising flux due to the aging of the stellar population, 
effectively highlight the difficulties of reproducing simultaneously all 
the observational constraints by means of traditional photoionisation models.

Since the essence of this problem is finding a way of reducing 
the predicted dispersion,
one would na\"\i vely expect that the problem would worsen
when the IMF sampling effects are taken into account:
that is, that the positions occupied in diagnostic diagrams 
by model nebulae would show a larger dispersion
as a consequence of taking into account
the dispersion in the ionising flux
predicted by stochastic models.
Instead, we show here that the intensities 
of the predicted flux at different energies
are strongly correlated, a fact implying
that no additional dispersion will result from
the incomplete sampling of the IMF.
 
The structure of this paper is the following: in Section~\ref{sec:sampling} 
we review briefly
the evaluation of sampling effects and the definition of the Lowest
Luminosity Limit.  The application of these concepts to the case of
ionising spectra is presented in Section~\ref{sec:methodology}.  
Our main results are explained in Section~\ref{sec:results},
and the conclusions are presented in Section~\ref{sec:conclusions}.

\section{Consequences of sampling effects in synthesis models}\label{sec:sampling}

In this section we describe how sampling
effects in the IMF are evaluated. 
Some of these results have either been presented in previous papers,
or can be found in books on statistics, e.g. \cite{KS77}: 
nevertheless, we have chosen to
summarize them here rather than in an appendix, because they are
essential for the forthcoming discussion.  Readers that are
already familiar with these tools 
may jump directly to Section~\ref{sec:methodology}.

\subsection{The Lowest Luminosity Limit}

Sampling effects produce a dispersion in the results
of synthesis models, and should therefore be taken into
account when the properties of real clusters
are interpreted by means of synthesis models.
The relevance of such effects depends on the size of the
system studied. 
A starting point to traduce this statement into quantitative terms
is the definition of a criterion, related to the cluster size,
allowing one to establish whether sampling effects
in a particular cluster
are playing a fundamental role or not.
This alternative in turn would condition 
the mode of application of synthesis models:
clusters small according to this criterion should necessarily
be modeled with full consideration of sampling effects,
whereas larger clusters could be modeled
by means of traditional analytical methods.

A suitable criterion to classify systems in this way
is provided by the Lowest Luminosity Limit
\citep[or LLL:][]{CL03,CL03b}.
The LLL is defined as the luminosity of the 
most luminous individual star 
compatible with the physical properties (age, metallicity, etc.)
of the cluster,
and the criterion to follow
is that an observed cluster whose luminosity
is below the LLL cannot be modeled by means of a method 
that neglects sampling effects.
The logic underlying this definition is that
an analytical model computed to reproduce a cluster
below the LLL
necessarily includes fractions smaller than one 
of the relevant stars,
and therefore lack physical meaning. 
Note that the restriction imposed by this criterion
is quite loose, since it constrains only those
situations in which strong statistical effects are 
undoubtedly present; but important statistical effects may well 
be present even in clusters above the LLL.

Since the average luminosity $A$ of a cluster scales
linearly with the total mass ${\cal M}$,
the LLL is naturally associated to a lower mass limit ${\cal M}_A^{min}$,
which is the mass of a stellar system with a
completely sampled IMF and luminosity equal to the LLL.

The numerical value of the LLL depends on the isochrones 
and the atmosphere models used, 
does not depend on the IMF, and is only weakly dependent on
the star formation history; on the contrary, it depends strongly 
on the age and metallicity of the synthetic cluster.
Finally, the LLL obviously depends on the energy band
in which the luminosity is defined.
For a given luminosity band $A$ the LLL defines
an intrinsic limit below which a computation of the statistical 
dispersion associated to the incomplete sampling of the IMF
is unavoidable for a meaningful application of synthesis models.
Above the LLL,
it is possible to provide a rough estimation of the dispersion
and a guideline to determine the relevance of sampling effects: 
for example,
\cite{CL03} estimate that the dispersion in the results of synthesis models
for clusters with initial masses about $10\times {\cal M}_A^{min}$ is equal
or larger than 10\% in the optical and infrared bands. 
Below the LLL, a proper statistical formalism 
is required to obtain quantitative estimations
of the relative dispersion in the predicted quantities;
this topic will be covered in the next section.
However, note that below the LLL 
the predicted average values of observables that do not scale
linearly with ${\cal M}$, like logarithmic or rational functions of
luminosities (i.e., equivalent widths or colours) can be severely biased
with respect to actual observations \cite[see][]{CVG03},
so that even a sophisticated statistical formalism can fail
to produce meaningful predictions.

\subsection{Estimation of sampling effects}\label{sec:sampeffects}

As stated in Section~\ref{sec:introduction},
sampling effects in synthesis models can be directly estimated 
by Monte Carlo methods, or alternatively by a statistical
formalism applied to the results of analytical synthesis models.
A formalism of this kind, based on the original formalism proposed by
\cite{Buzz89}, has been developed in recent years by
\cite{CLC00,Cetal01,CVGLMH02}; and \cite{CVG03}.  The method can be briefly
summarised as follows. Let us assume that $w_i$ is the number of stars
within a given mass range, normalised to the total mass; this value is
given by the IMF and the star formation history. Assuming a Simple Stellar
Population model (SSP, i.e. a model with coeval star formation, also called
Instantaneous Burst model), $w_i$ is given by integrating the IMF over the
corresponding mass range. Each $w_i$ is approximately the mean value of a
Poisson distribution, and so the variance $\sigma_i^2$ of each $w_i$ is
equal to $w_i$.  This assumption is a good approximation if the mass
interval that defines the $w_i$ values is narrow enough: see 
\citet{CVGLMH02} for a complete discussion of this topic.

Given an observable property $a_i$ of a given star $i$, its contribution to
the average normalised integrated property $\mu_A^{ssp}$ obtained by the
synthesis model is given by $w_i a_i$, with a variance $\sigma_i^2 a_i^2 =
w_i a_i^2$. The total variance of the average observable $\mu_A^{ssp}$ is
the sum of all the variances. Hence the relative dispersion is simply

\begin{equation}
\frac{\sigma_{A}^{ssp}}{\mu_A^{ssp}} \; = \; 
\frac{(\sum_{i} w_i a_i^2)^{1/2}}{\sum_{i} w_i a_i}  \; = \; 
\frac{1}{\sqrt{{\cal{N}}_A^{ssp}}} \, ,
\label{eq:neff}
\end{equation}

\noindent where the last term gives the definition of ${\cal{N}}_A$ introduced by
\citet{Buzz89}. Since $w_i$ scales with the initial mass ${\cal M}$ transformed into
stars in the cluster, the effective number of stars contributing
to observable $A$ also scales with the initial mass. This implies that, if
the ${\cal M}$ value of an observed cluster is known, its predicted
average $A$ value is $\mu_A = \mu_A^{ssp} \times {\cal M}$, and the
dispersion around $\mu_A$ is given by ${\cal N}_A = {\cal N}_A^{ssp}
\times {\cal M}$.

In addition, given another property $b_i$, with contribution to
the integrated average property $\mu_B^{ssp}$ given by $w_i b_i$ and 
variance $(\sigma_B^2)^{ssp}$, it is also possible to obtain the linear
correlation coefficient between the observables $A$ and $B$:

\begin{eqnarray}
\rho(A,B)&=&\frac{\mathrm{cov}(A,B)}{\sigma_A \sigma_B}=
\frac{\mathrm{cov}(A,B)^{ssp}}{\sigma_A^{ssp} \sigma_B^{ssp}}\nonumber \\
&=&\frac{\sum_{i} w_i a_i b_i}{\sqrt{(\sum_{i} w_i a_i^2)(\sum_{i} w_i
b_i^2)}} \, ,
\label{eq:rho}
\end{eqnarray}

\noindent where $\mathrm{cov}(A,B)$ is the covariance.
 Although $\mathrm{cov}(A,B)$ scales with ${\cal M}$,
$\rho(A,B)$ is independent of it. 
The correlation coefficient can also
 be obtained by a linear regression of $A$ vs. $B$: 
a value of $\rho(A,B)=1$ implies that
the two quantities are linearly dependent. 
In terms of a hypothetical diagnostic diagram of $A$ vs. $B$,
$\rho(A,B)\sim 1$ implies that  the data
points follow a narrow sequence.

\section{Methodology}\label{sec:methodology}

\begin{figure}
  \resizebox{\hsize}{!}{\includegraphics[width=7cm]{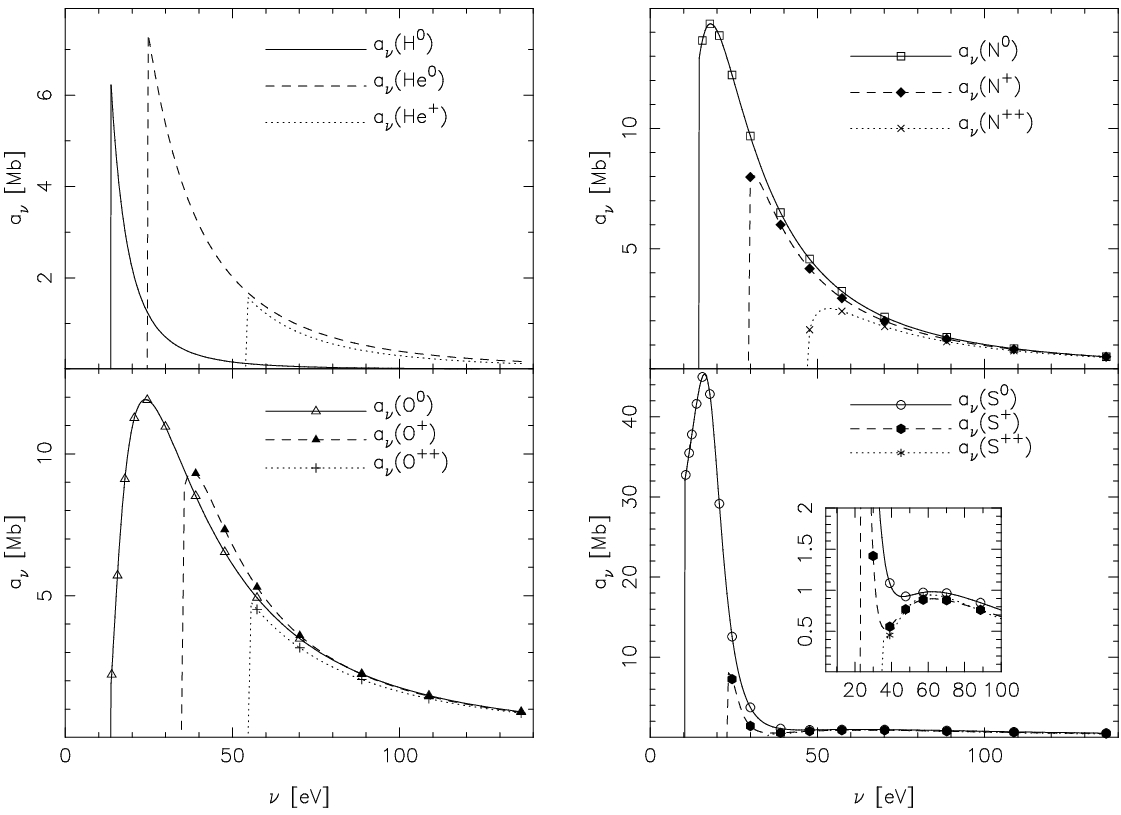}}
  \caption[]{Photoionisation cross sections $a_\nu(X^{i})$ for the ions
considered in this paper, as obtained from the routine {\tt phfit.f} in
  the photoionisation code {\tt Cloudy} \citep{ferland}.}
\label{fig:CS}
\end{figure}

The simplest way to characterize the ionising flux of a
stellar population
is to consider both its intensity (in number of photons emitted) and
its shape (in terms of 
``effective rate of ionising photons'' 
defined
below).
The emission rate of photons capable of ionising a particular ion
($Q(\mathrm{H}^{0})$, $Q(\mathrm{He}^+)$, etc.) with ionisation 
threshold frequency $\nu_0$ is given by:

\begin{equation}
Q(X^{i})=\int_{\nu_0}^\infty \frac{L_\nu}{h\nu} \; d\nu \, .
\end{equation}

As for the shape of the ionising flux, it is important to note that not all
ionising photons are equivalent since the photoionisation cross section
depends strongly on $\nu$ (see Fig. \ref{fig:CS}). The effectiveness of
photons in ionising a ion $X^{i}$ is measured
by the balance between $X^{i}$ and $X^{i+1}$
at a given point in the nebula:

\begin{equation}
\frac{N(X^{i+1}) N_e}{N(X^{i})}= \frac{ \int_{\nu_i}^\infty ({4 \pi
J_\nu}/{h\nu}) a_{\nu}(X^{i}) d\nu} {\alpha_G(X^{i},T)} \, ,
\label{ion0}
\end{equation}

\noindent where $N(X^{i+1})$, $N(X^{i})$, and $N_e$ are the number densities
of ions $X^{i+1}$, $X^{i}$, and electrons respectively, $\alpha_G(X^{i},T)$
is the recombination coefficient of the ground level of $X^{i+1}$ to all
levels of $X^i$, $a_{\nu}(X^{i})$ is the absorption cross section of the
ion $X^{i}$ from its ground level with threshold $\nu_i$, and $4 \pi
J_\nu$ is the local flux density \citep{O94}.  
Since we are ultimately interested in determining the emission
spectra, which in turn depend on the relative ionic abundances,
it seems physically meaningful
to weigh the emission rate of photons capable to ionise a particular ion
by the photoionisation cross section for that ion, in order to describe 
the luminosity actually seen by each ion;
The resulting quantity, which we call the 
``effective rate of ionising photons'' 
is defined by the expression:

\begin{equation}
Q_{a_\nu}(X^{i})= \int_{\nu_i}^\infty \frac{L_\nu}{h\nu} a_{\nu}(X^{i}) \, d\nu
\label{ion} \, .
\end{equation}

This quantity is obviously not an observable, but it will help to perform
first-order estimations of the sampling effects on the ionising flux, via 
Eq. \ref{ion0}. 
We have computed $Q_{a_\nu}(X^{i})$ for 
H$^{0}$, He$^{0}$, He$^{+}$, N$^{0}$, N$^{+}$, N$^{++}$, O$^{0}$,
O$^{+}$, O$^{++}$, S$^{0}$, S$^{+}$, S$^{++}$, C$^{0}$, Ne$^{0}$, Ne$^{+}$,
Ar$^{0}$, Ar$^{+}$, and Ar$^{++}$, although in this work the Ar, Ne, and C
ions 
will not be discussed\footnote{The complete set of results can
be found at the CMHK web pages at {\tt http://laeff.inta.es/users/mcs/SED/.}}.
The corresponding $a_\nu$ values have been
obtained from the subroutine {\tt phfit.f} used for this purpose in the
photoionisation code {\tt Cloudy} \citep{ferland}, and are shown in
Fig. \ref{fig:CS}. 
It is important to recall that since the recombination lines of $X^i$ arise
from the recombination of $X^{i+1}$, they depend on $Q_{a_\nu}(X^{i})$
(e.g., $Q_{a_\nu}(\mathrm{H}^{0})$ is related to the \ion{H}{i} lines).
On the other hand, 
since forbidden lines of $X^{i}$ arise from $X^{i}$ levels, 
they depend on $Q_{a_\nu}(X^{i-1})$
(e.g.  $Q_{a_\nu}(\mathrm{O}^{+})$ is related to the [\ion{O}{iii}]
lines, and not to the [\ion{O}{ii}] lines).

\section{Results}\label{sec:results}

In this section we will describe our results. 
All the computations assume an instantaneous burst of star formation
and a \cite{sal55} IMF in the mass range 0.1 to 120 M$_\odot$.

To assess the possible systematic effects associated with the choice
of stellar tracks, model atmospheres, interpolation schemes, etc, 
we have used two different evolutionary synthesis codes:

\begin{itemize}

\item The code by \cite{CMHK02} (hereinafter CMHK), 
which adopts evolutionary tracks with standard mass-loss rates
\citep{Schetal92,Schetal93,Schetal93b,Charetal93} and the model atmospheres
by \cite{CoStar} ({\sc CoStar}) for main sequence hot
stars
more massive than 20 M$_\odot$,
by \cite{Schmetal92} for WR stars, 
and by \cite{kur} for the remaining stars.

\item A modified version of {\sc Starburst99} \citep{SB99},
where the main modification is the computation of the $Q_{a_\nu}(X^{i})$
values and the corresponding minimum masses.
The use of {\sc Starburst99} is only for comparison purposes,
since we are interested in studying the impact of the
new atmosphere models by \cite{linda} for WR
and hot stars, implemented in this code. 
Note that {\sc Starburst99} assumes
Blackbody spectra for stars with $\log T_{eff} > 4.778$ or $\log T_{eff} <
3.3$ except for WR stars (defined by the stellar temperature, $\log T_{*} >
4.4$, the hydrogen surface abundance $X_s < 0.4$, and the minimum mass for
WR formation). The stellar tracks used have
high mass-loss rates \citep{Mey94}.

\end{itemize}

In both codes, the spectrum associated to a given star is given by the
closest model atmosphere found either in the $\log g - T_{eff}$ plane or in
the $g - T_{eff}$. For the computation of the isochrones, both codes follow
the prescriptions described in \cite{Cetal01}.

\begin{figure*}
  \resizebox{\hsize}{!}{\includegraphics[width=16cm]{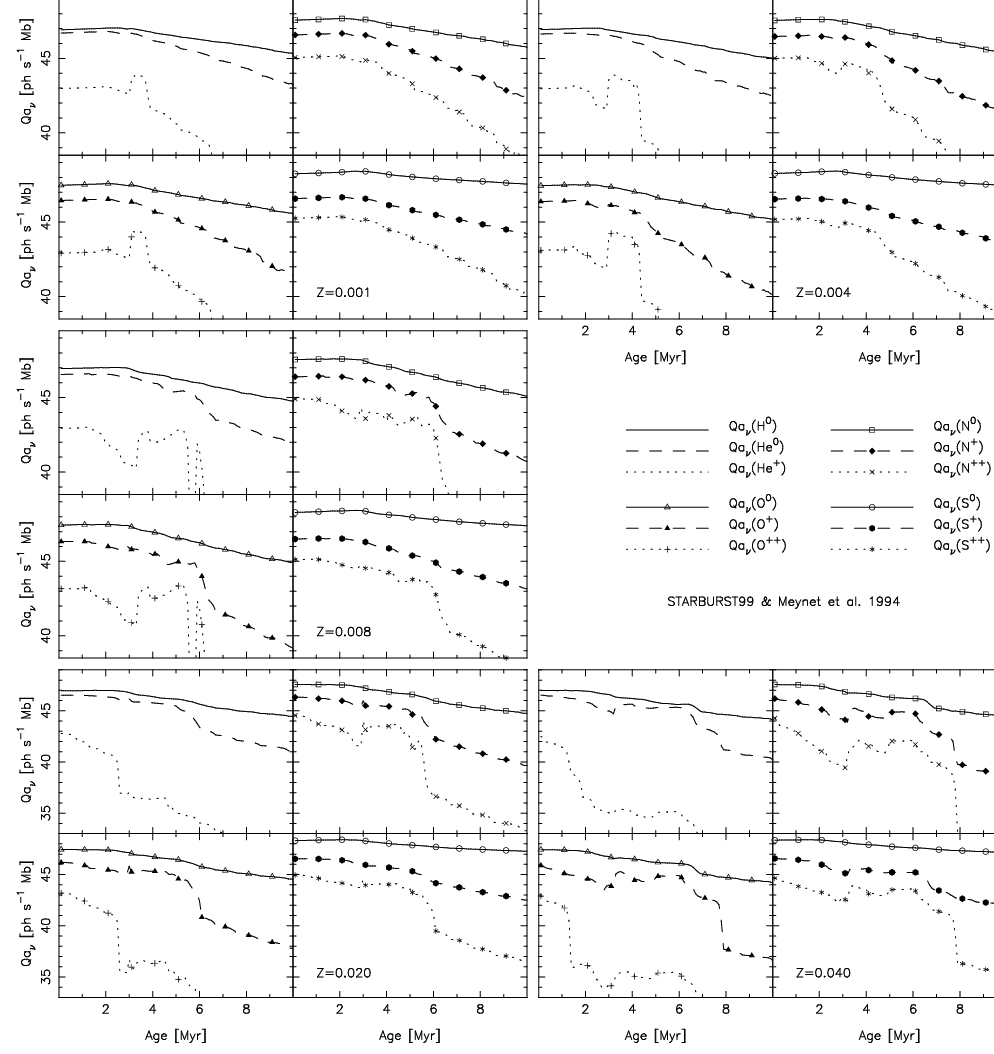}}
  \caption[]{Evolution of the cross-section-weighted number of 
  ionising photons $Q_{a_\nu}(X^{i})$ as a function of time for different
  metallicities and ions, using {\sc
  Starburst99}  (stellar tracks with enhanced mass-loss rates, and
  model atmospheres by \cite{linda}).  The values are
  normalized to 1 M$_\odot$ for a cluster with a Salpeter IMF in the mass
  range 0.1 -- 120 M$_\odot$. 
 The five main panels correspond to the
metallicities $Z$=0.001 (top-left), $Z$=0.004 (top-right), $Z$=0.008
(middle), $Z$=0.020 (solar, bottom-left) and $Z$=0.040 (bottom-right).  Each main panel is further subdivided into
four sub-panels: the top-left sub-panel corresponds to
H$^{0}$, He$^{0}$, and H$^{+}$; the top-right sub-panel
to N$^{0}$, N$^{+}$, and N$^{++}$; the
bottom-left sub-panel to O$^{0}$, O$^{+}$, and O$^{++}$; 
and the bottom-right sub-panel to S$^{0}$, S$^{+}$, and S$^{++}$.}
\label{fig:Qsb99}
\end{figure*}

\begin{figure*}
  \resizebox{\hsize}{!}{\includegraphics[width=16cm]{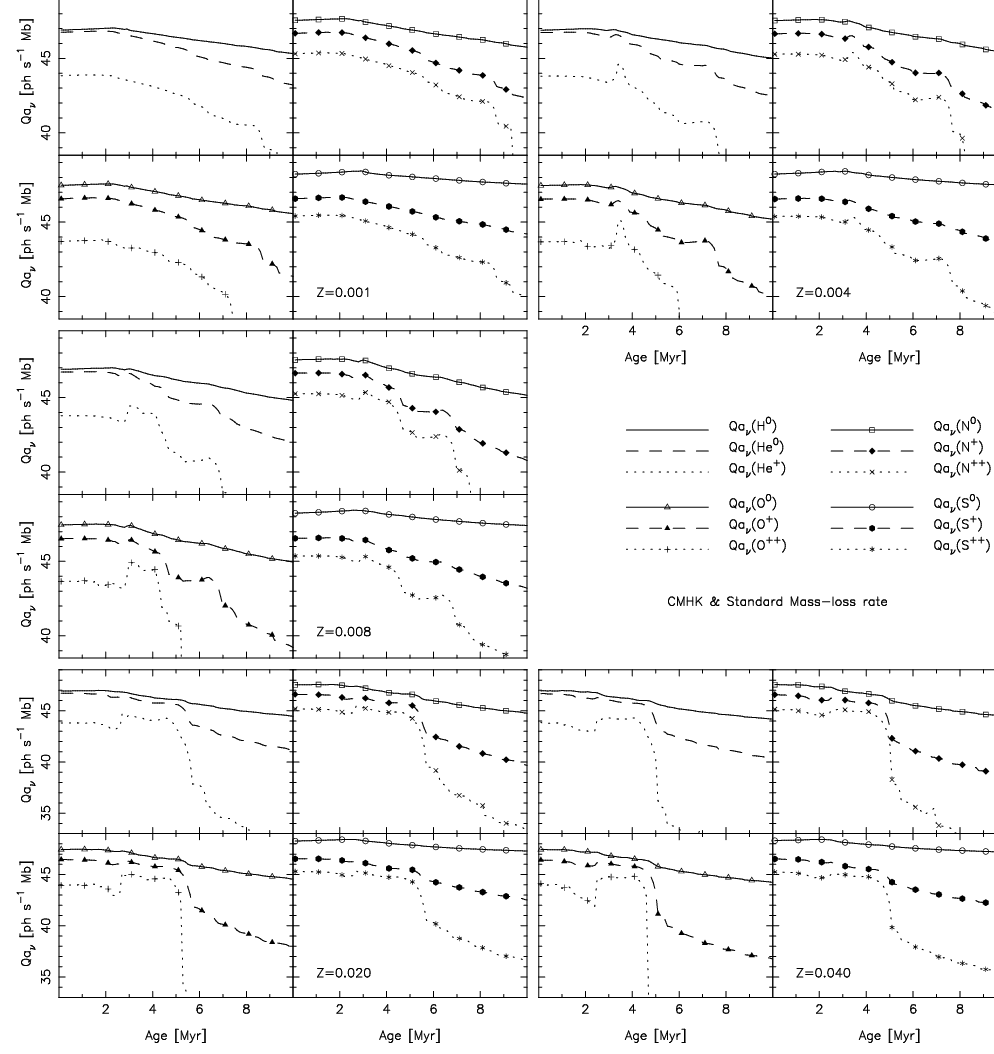}}
  \caption[]{Evolution of the cross-section-weighted number of 
  ionising photons $Q_{a_\nu}(X^{i})$ as a function of time for different
  metallicities and ions, using the CMHK
  code (stellar tracks with standard mass loss rates).
Panel division and symbols like in Fig. \ref{fig:Qsb99}.
}
\label{fig:Qsed}
\end{figure*}

\subsection{Effective rate of ionising photons $Q_{a_\nu}(X^{i})$}

The time evolution of $Q_{a_\nu}(X^{i})$ predicted by {\sc Starburst99} 
and the CMHK code is shown in Fig. \ref{fig:Qsb99} and \ref{fig:Qsed} 
respectively,
where the five main panels correspond to the
metallicities indicated\footnote{Note the different scale along the $Q_{a_\nu}$-axis
for $Z$=0.020 and $Z$=0.040.}.  Each main panel is further subdivided into
four sub-panels, each showing the results for a different set of ions, 
coded according to
the lines and symbols indicated. The top-left panel corresponds to
H$^{0}$, He$^{0}$, and H$^{+}$; the top-right panel
to N$^{0}$, N$^{+}$, and N$^{++}$; the
bottom-left panel to O$^{0}$, O$^{+}$, and O$^{++}$;
and the bottom-right panel to S$^{0}$, S$^{+}$, and S$^{++}$.

There are striking differences between the $Q_{a_\nu}(X^{i})$
behaviour predicted by the two codes,
a fact that emphasises the importance of the models' input
-- evolutionary tracks and atmosphere models -- as a source of
systematic effects.  
An example is the difference in the bumps of 
the $Q_{a_\nu}({\rm He}^{+})$ curves:
these bumps reveal the life cycle of WR stars, 
which provide a substantial amount of hard photons.
The low $Z$ models computed with {\sc Starburst99}
show such bumps at about 3 Myr,
whereas the CMHK models do not; 
the difference is mainly due 
to the different evolutionary tracks,
because enhanced mass-loss rates produce larger numbers of WR stars.
A further difference is the larger $Q_{a_\nu}({\rm He}^{+})$ value 
at high $Z$ in the CMHK code as compared to
{\sc Starburst99}. This difference can be ascribed
to differences in the atmosphere models implemented in the two 
codes:
in general, the CMHK code produces a harder ionising
flux due to the use of the {\sc CoStar} and
\cite{Schmetal92} atmosphere models.
This effect is larger for high metallicities during
the WR phase.  The impact of different atmosphere models in synthesis codes
has been recently discussed by \cite{linda}, and we refer to that paper for
further details.

\begin{figure}
  \resizebox{\hsize}{!}{\includegraphics[width=7cm]{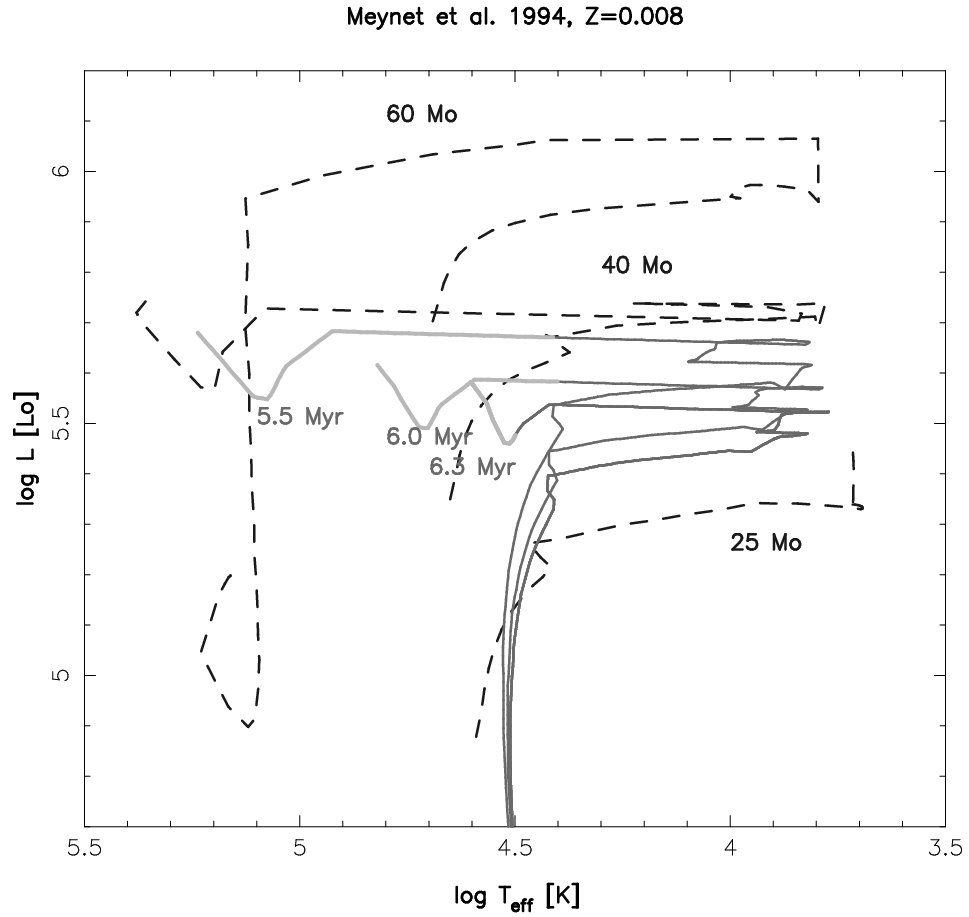}}
  \caption[]{Evolutionary tracks (dashed line) and isochrones (solid lines)
from {\sc Starburst99} and \cite{Mey94}.}
\label{fig:iso}
\end{figure}

There are two other interesting features in 
Fig. \ref{fig:Qsb99} and \ref{fig:Qsed} 
that deserve discussion:

\begin{enumerate}

\item Although the behaviours of $Q_{a_\nu}(\mathrm{He}^{+})$ and
$Q_{a_\nu}(\mathrm{O}^{++})$ are remarkably similar, 
as a consequence of their
ionisation edges being almost identical (54.418 eV for He$^{+}$ and 54.936 eV
for O$^{++}$), there are slight differences between them
due to the greater sensitivity to low energy photons 
of $a_\nu(\mathrm{He}^{+})$ with respect to $a_\nu(\mathrm{O}^{++})$.  
It follows that $Q_{a_\nu}(\mathrm{O}^{++})$
decreases more rapidly than $Q_{a_\nu}(\mathrm{He}^{+})$ 
when the hard ionising flux decreases, that is for ages $t\grsim 4$ Myr.

\item The tails of the $Q_{a_\nu}(\mathrm{He}^{+})$ and
$Q_{a_\nu}(\mathrm{O}^{++})$ curves during the post-WR
phases show an interesting behaviour in
both sets of results.
For example, at $Z$=0.008 
there is a $Q_{a_\nu}(\mathrm{He}^{+})$ peak around 6 Myr
in the case of stellar tracks 
with enhanced mass-loss rates
(Fig. \ref{fig:Qsb99}),
and a peak around 6.5 Myr in the case of stellar tracks 
with standard mass-loss rates (Fig. \ref{fig:Qsed}).
Yet, although the bumps in these curves reveal the presence 
of WR stars in the stellar population,
the WR phase ends at 5.9 Myr in the 
case of enhanced mass-loss rates,
and at 4.7 Myr in the case of standard mass-loss rates.
The explanation of this apparent contradiction 
lies in the track interpolation technique used to
obtain the isochrones: even though it is assumed that the evolutionary
tracks follow a continuous sequence, there is a discontinuity in the
stellar tracks between WR and non-WR phases.  The situation is illustrated
in Figure \ref{fig:iso}, where isochrones for 4.5, 6 and 6.3 Myr obtained
from {\sc Starburst99} are compared with the
evolutionary tracks for 60, 40 and 25 M$_\odot$ with enhanced mass-loss
rates. Note that some stars in the 6 Myr isochrone (where no WR stars 
should be present) reach $\log T_{eff}$ larger than 4.4, 
and they are therefore (wrongly) assigned to the WR type. 
A similar problems occurs in the CMHK code, 
and in general in all synthesis codes, since continuity
in the tracks is implicitly assumed \cite[see, however, an alternative
approach presented in][]{CMH94}. The isochrones are mathematically
consistent, but this is not a guarantee that they are physically
consistent. This problem is amplified in the tracks with enhanced
mass-loss rates since adjacent tracks are far more different than the
corresponding tracks with standard mass-loss rates.  This may superficially
be considered a technical detail \citep{MaeGal2}, but since it may produce
unphysical results, it is worth investigating it in detail (Cervi\~no 2003,
in preparation).

\end{enumerate}

\subsection{Minimum masses ${\cal M}^{min}[Q_{a_\nu}(X^{i})]$}

\begin{figure*}
  \resizebox{\hsize}{!}{\includegraphics[width=16cm]{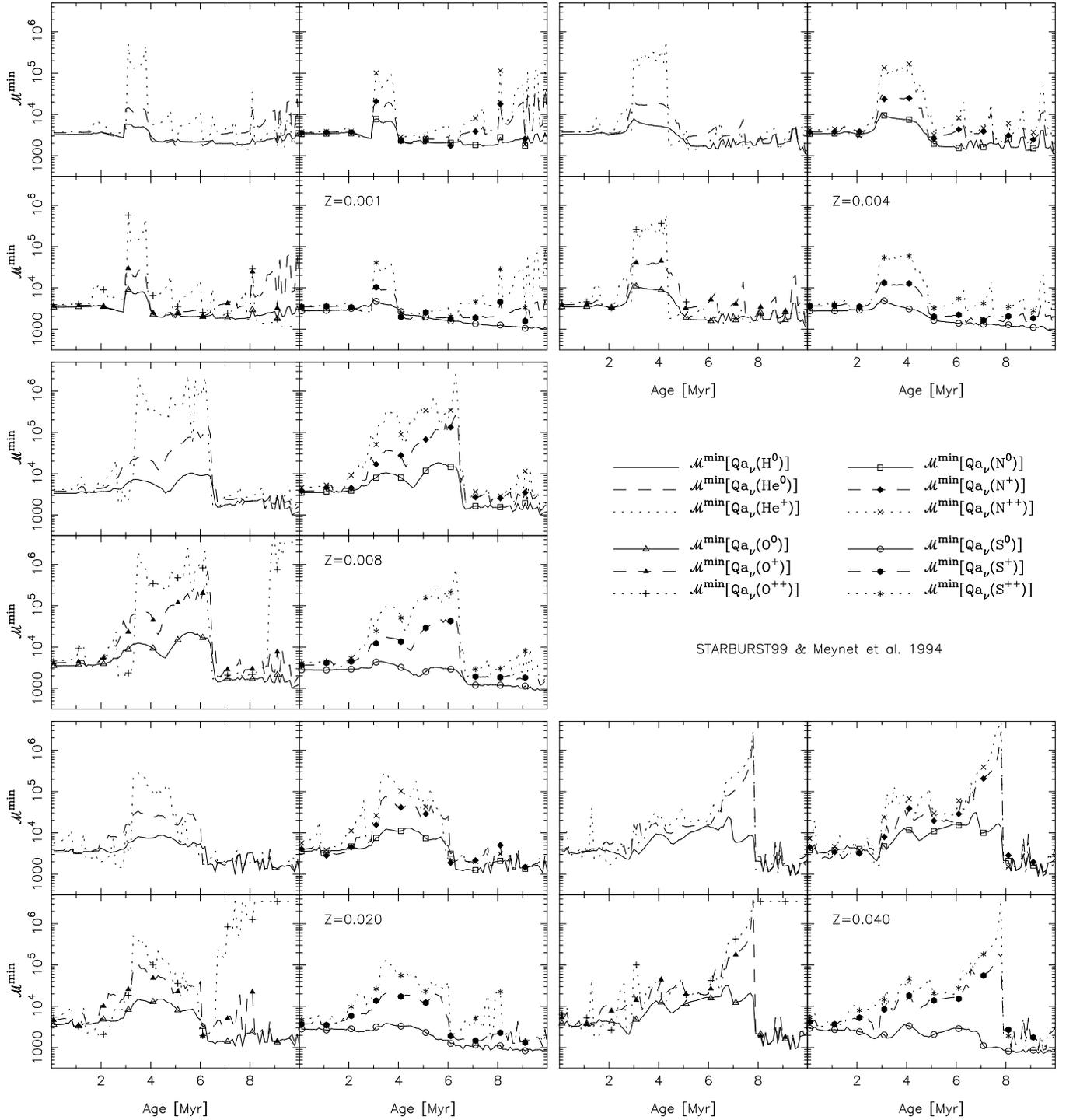}}
  \caption[]{${\cal{M}}^{min}[Q_{a_\nu}(X^{i})]$ values as a function of
  time for different ions and metallicities using {\sc Starburst99} with
  stellar tracks with enhanced mass-loss rates, and atmosphere models
  by \cite{linda}.  The values are normalised to 1 M$_\odot$
  for a cluster with a Salpeter IMF in the mass range 0.1 -- 120
  M$_\odot$. 
Panel division and symbols like in Fig. \ref{fig:Qsb99}.
}
\label{fig:Mminsb99}
\end{figure*}

\begin{figure*}
  \resizebox{\hsize}{!}{\includegraphics[width=16cm]{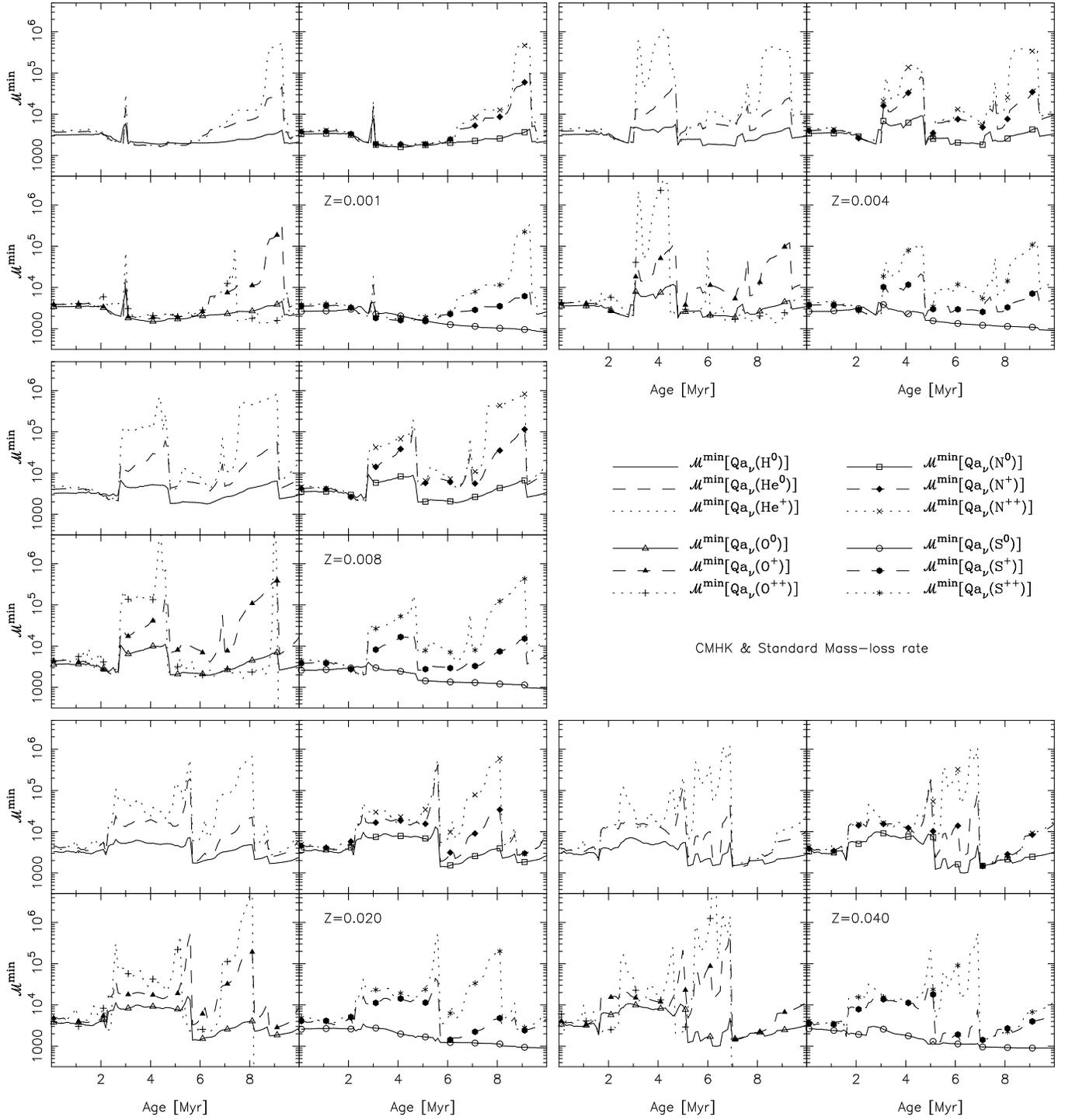}}
  \caption[]{Same as in Fig.~\ref{fig:Mminsb99}, but using the
  CMHK code and standard mass-loss rate tracks.
Panel division and symbols like in Fig. \ref{fig:Qsb99}.
}
\label{fig:Mminsed}
\end{figure*}

In analogy to our definition of the minimum mass associated to the LLL, 
we will define in the following a minimum mass associated 
to the effective rate of ionising photons, ${\cal M}^{min}[Q_{a_\nu}(X^{i})]$,
which is defined as the mass of a stellar system with a
completely sampled IMF and total effective rate of ionising photons
equal to the maximum of all the effective rates
provided by the individual stars in the system.

The evolution of ${\cal M}^{min}[Q_{a_\nu}(X^{i})]$ for the ions selected
before, computed with {\sc Starburst99} and the CMHK code, are shown in
Fig.~\ref{fig:Mminsb99} and \ref{fig:Mminsed} respectively. The
organisation of the figures is the same as in Fig.~\ref{fig:Qsb99}.
Perhaps the most striking feature of these results is the huge mass range
covered, which spans nearly three orders of magnitude, from about $3\times
10^3$ to more than 10$^6$ M$_\odot$ for some of the important ions,
independently of the code used.  As expected, the larger the ionising
potential, the larger the minimum mass ${\cal{M}}^{min}[Q_{a_\nu}(X^{i})]$,
since incomplete sampling affect mainly the most massive stars, which are
those contribute most to this property.  Likewise, S$^{0}$ has the lowest
${\cal{M}}^{min}[Q_{a_\nu}(X^{i})]$ value: the ionising potential of
S$^{0}$ is only 10.36 eV, so the effective rate of ionising photons for
this ion is scarcely affected by sampling effects.  On the other hand,
He$^{+}$ is the ion more affected by sampling effects: this result confirms
that \ion{He}{ii} lines cannot be used as a strong constrain on the
properties of stellar clusters, as suggested by \citet{Lal03}.

The most prominent peak in ${\cal{M}}^{min}[Q_{a_\nu}(X^{i})]$
corresponds to the WR phase of the cluster. In the case of {\sc
Starburst99}, there is a tendency to obtain larger
${\cal{M}}^{min}[Q_{a_\nu}(X^{i})]$ values during the WR-phase at
intermediate metallicities ($Z$=0.008), and an asymmetric ``U-shaped''
curve at low metallicities. We have tested this behaviour obtaining the
corresponding ${\cal{M}}^{min}[Q_{a_\nu}(X^{i})]$ with the CMHK code using
the evolutionary tracks by \cite{Mey94}\footnote{Which results are also
available at the CMHK web pages.}, and the ``U-shaped'' behaviour
remains. However, ${\cal{M}}^{min}[Q_{a_\nu}(X^{i})]$ tends to be lower for
larger metallicities.  For the case of the CMHK code, the effect is
explained easily: since the WR atmosphere models are the same for all the
metallicities, the larger the metallicity, the larger the mass-loss rate,
the lower the luminosity, and the lower
${\cal{M}}^{min}[Q_{a_\nu}(X^{i})]$.  In the case of {\sc Starburst99} the
effect of the metallicity-dependent WR atmosphere models must also be taken
into account together with the variation of the luminosity with
metallicity. In fact, examining the ionising fluxes quoted for the WR stars
in tables 3 and 4 of \cite{linda}, it is found that the flux of ionising
photons is not a monotonic function of the metallicity: the largest 
values for the rate of ionising photons correspond to some of the WR
models at $Z$=0.008 ($Z$=0.4 $Z_\odot$ in their paper).
This would be the origin of the
larger ${\cal{M}}^{min}[Q_{a_\nu}(X^{i})]$ value at this metallicity. This
also shows the impact of the use of metallicity-dependent atmosphere models
for WR stars.

The ``U-shaped'' behaviour during the WR-phase is due to the evolutionary
tracks used: since the lower the number of stars that dominate the
luminosity, the larger the ${\cal{M}}^{min}[Q_{a_\nu}(X^{i})]$ value, its
value depends on the relative rate of ionising photons emitted by WR stars
with respect to the other hot stars in the synthetic cluster.  In the
tracks by \cite{Mey94}, the WR stars at the beginning of the WR phase are
also the most luminous stars in the cluster (see, for example, the 60
M$_\odot$ track in Fig. \ref{fig:iso}) with the harder ionising flux: this
fact produces the first peak in ${\cal{M}}^{min}[Q_{a_\nu}(X^{i})]$. When
the system evolves, the luminosity of WR stars decreases and 
the other non-WR, hot stars give only a 
small contribution (see, for example, the 40
M$_\odot$ track in Fig. \ref{fig:iso}), leading to the decrease in
${\cal{M}}^{min}[Q_{a_\nu}(X^{i})]$. 
However, since the non-WR stars become cooler as the age increases, 
at the end of the WR-phase
WR stars become again the almost only source of ionising flux, 
producing the secondary peak. Note also 
that, for $Z$=0.040, the end of the WR-phase 
is marked by the highest peak in the curve: 
this behaviour is due to the longer duration of the WR-phase at this
metallicity and to the dependency of the lifetime of massive stars with
mass, which in these tracks is not monotonic: stars with initial mass of 85
and 120 M$_\odot$ stars end their evolution as WR at 5 and 8 Myrs
respectively. As a consequence, in such an age range and at this
metallicity, the ionising flux is dominated by the less populated tail of
the IMF, so that sampling effects dominate and a larger
${\cal{M}}^{min}[Q_{a_\nu}(X^{i})]$ value is obtained.

The same general explanations also apply
to the case of the CMHK results with standard mass-loss rate.
However, for these tracks WR and non-WR
stars have more similar luminosities and the first peak does not appear (the
curve is ``J-shaped''). In this case the maximum in
${\cal{M}}^{min}[Q_{a_\nu}(X^{i})]$ occurs at $Z$=0.004, because the
WR phase is very short and almost absent at $Z$=0.001. In fact, at this
metallicity, there no star ends its evolution as a WR; the
WR-phase only appears in the {\it middle} of the 120
M$_\odot$ track, which ends as a supergiant.  

There are also several peaks in ${\cal{M}}^{min}[Q_{a_\nu}(X^{i})]$ at the
oldest ages considered in these plots.  These features are an artefact of
the way in which model atmospheres are assigned to stars in the synthetic
HR diagram: since the smooth evolutionary path of each star in the HR
diagram is coupled to a discrete ensemble of atmosphere models, a peak is
produced when stars suddenly switch from one atmosphere model to the next.
In the case of {\sc Starburst99}, the effect is not too strong because this
code uses a finer atmosphere grid. In the case of the CMHK code, the switch
from the {\sc CoStar} to the \cite{kur} atmosphere models for stars in the
main sequence around $M=20$ M$_\odot$ produces
the additional bumps in ${\cal{M}}^{min}[Q_{a_\nu}(X^{i})]$ with the
most prominent ones ending at $t \approx$ 9.4 Myr for $Z$=0.001 and
$Z$=0.004, $t \approx$ 9 Myr for $Z$=0.008, $t \approx$ 8.1 Myr for $Z$=0.020,
and $t \approx$ 7 Myr for $Z$=0.040, corresponding to the turn-off age 
of a 20 M$_\odot$ star (point 13 in the tracks), the phase at which 
{\sc CoStar} models are no longer used. Additional small bumps are found
at the turn-off age of a 25 M$_\odot$. We
want to remark again that these bumps are model artifacts, 
and that they would not appear
if the atmosphere models were interpolated, instead 
of being computed by means of 
the ``closest atmosphere model'' approach.  

Note that ${\cal M}^{min}[Q_{a_\nu}(X^{i})]$ corresponds to the 
amount of gas transformed into stars, and not to the total mass of the cluster 
(stars plus gas, that is dynamical mass, which would be approximately 
a factor of ten larger), 
and that a Salpeter IMF in the mass range 0.1 to 120 M$_\odot$ has
been assumed.  Table \ref{tab:IMF} gives conversion factors for different
mass limits and also the numbers of stars for a Salpeter IMF.  The results
of {\sc Starburst99} are usually normalised to $10^6$ M$_\odot$ in
the mass range 1--120 M$_\odot$, while CMHK normalise the results 
to 1 M$_\odot$ in the mass range 2--120 M$_\odot$.

\begin{table*}
\begin{center}
\begin{tabular}{lcccccc}
\hline
 & \multicolumn{6}{c}{IMF mass range}\\
\cline{2-7}\\ 
 Quantity     &0.1-120 & 1-120  & 2-120  & 8-120  & 25-120 & 40-120 \\
\hline
Mass into stars &   1    & 0.3962 & 0.2912 & 0.1442 & 0.0668 & 0.0428 \\
Number of stars & 2.8290 & 0.1262 & 0.0494 & 0.0074 & 0.0014 & 0.0007 \\
\hline
\end{tabular}
\end{center}
\caption[]{Conversion factors for cluster masses and numbers of stars for
different IMF mass ranges assuming a Salpeter slope.}
\label{tab:IMF}
\end{table*}

Using the values in Table \ref{tab:IMF}, a value of
${\cal{M}}^{min}[Q_{a_\nu}(X^{i})]$ around $3\times 10^3$ implies 
that there are, on average, two stars in the mass range 40--120 M$_\odot$, 
or 21 stars in
the mass range 8--120 M$_\odot$. Since the most massive stars are also the
most luminous ones, the ionisation of the cluster is accounted for by a few
stars. The presence of a few WR stars, with a harder ionising flux,
produces the increases in ${\cal{M}}^{min}[Q_{a_\nu}(X^{i})]$, especially
for ions with higher energy edges. A straightforward conclusion follows
from this result: the ionisation structure for clusters with ${\cal{M}}$
close to ${\cal{M}}^{min}[Q_{a_\nu}(X^{i})]$ would be better reproduced by
the ionising spectrum of a single star rather than by the synthetic
spectrum produced by a synthesis code,
which always includes an ensemble of stars.

\begin{figure*}
  \resizebox{\hsize}{!}{\includegraphics[width=16cm]{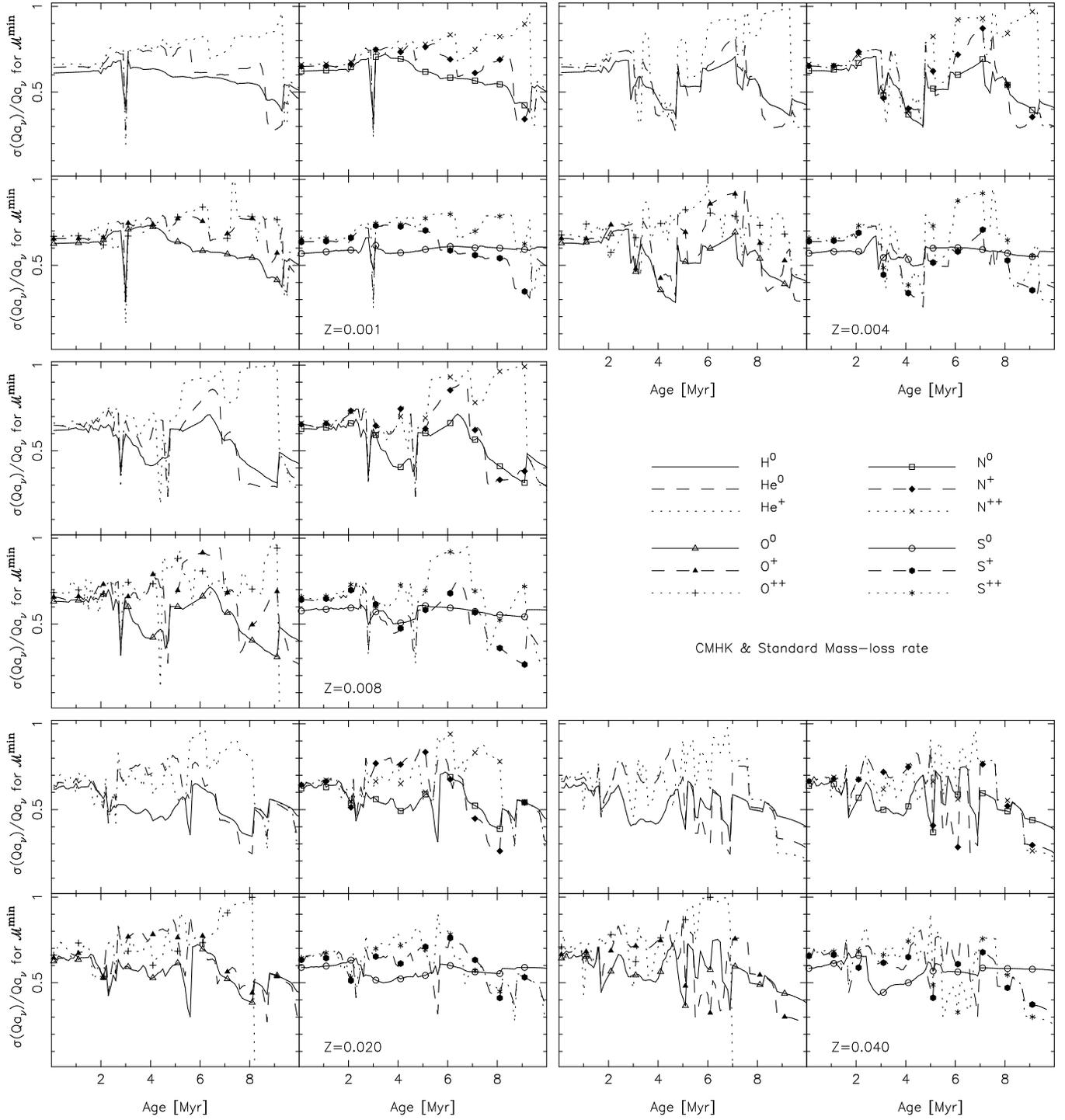}}
  \caption[]{Relative dispersion 
 $\sigma(Q_{a_\nu})/Q_{a_\nu}$  as a function of time for different
  metallicities and ions.
Panel division and symbols like in Fig. \ref{fig:Qsb99}.
}
\label{fig:NM}
\end{figure*}

It is necessary to recall, in this context, that
${\cal{M}}^{min}[Q_{a_\nu}(X^{i})]$ is, in fact, a rather restrictive 
lower limit for the use of integrated ionising spectra obtained by
synthesis models. A better way to estimate when a composite spectrum begins
to be a better approximation is given by
the relative dispersion in $Q_{a_\nu}(X^{i})$
corresponding to ${\cal{M}}={\cal{M}}^{min}[Q_{a_\nu}(X^{i})]$,
which is shown in Fig.~\ref{fig:NM}.
As expected, the dispersion varies strongly with the age, metallicity
and ion under consideration, but, in general, it has a value around
60\% (or, equivalently, ${\cal{N}}\sim 3$).  This means that in order
to obtain a dispersion smaller than 10\% (${\cal{N}}=100$), 
clusters with initial masses
${\cal{M}} \grsim 33\times{\cal{M}}^{min}[Q_{a_\nu}(X^{i})]$
are needed. This is a good lower limit
to ensure that the use of the composite ionising spectrum obtained from
synthesis models is appropriate. For clusters with masses in the range
[${\cal{M}}^{min}$, $33\times{\cal{M}}^{min}$] the situation is fuzzier, 
and any result obtained from codes which use a fully-sampled
IMF must be taken with extreme caution. The only proper alternative
solution in this case is to use Monte Carlo simulations.

\subsection{Correlations between ionising bands}

\begin{figure*}
  \resizebox{\hsize}{!}{\includegraphics[width=16cm]{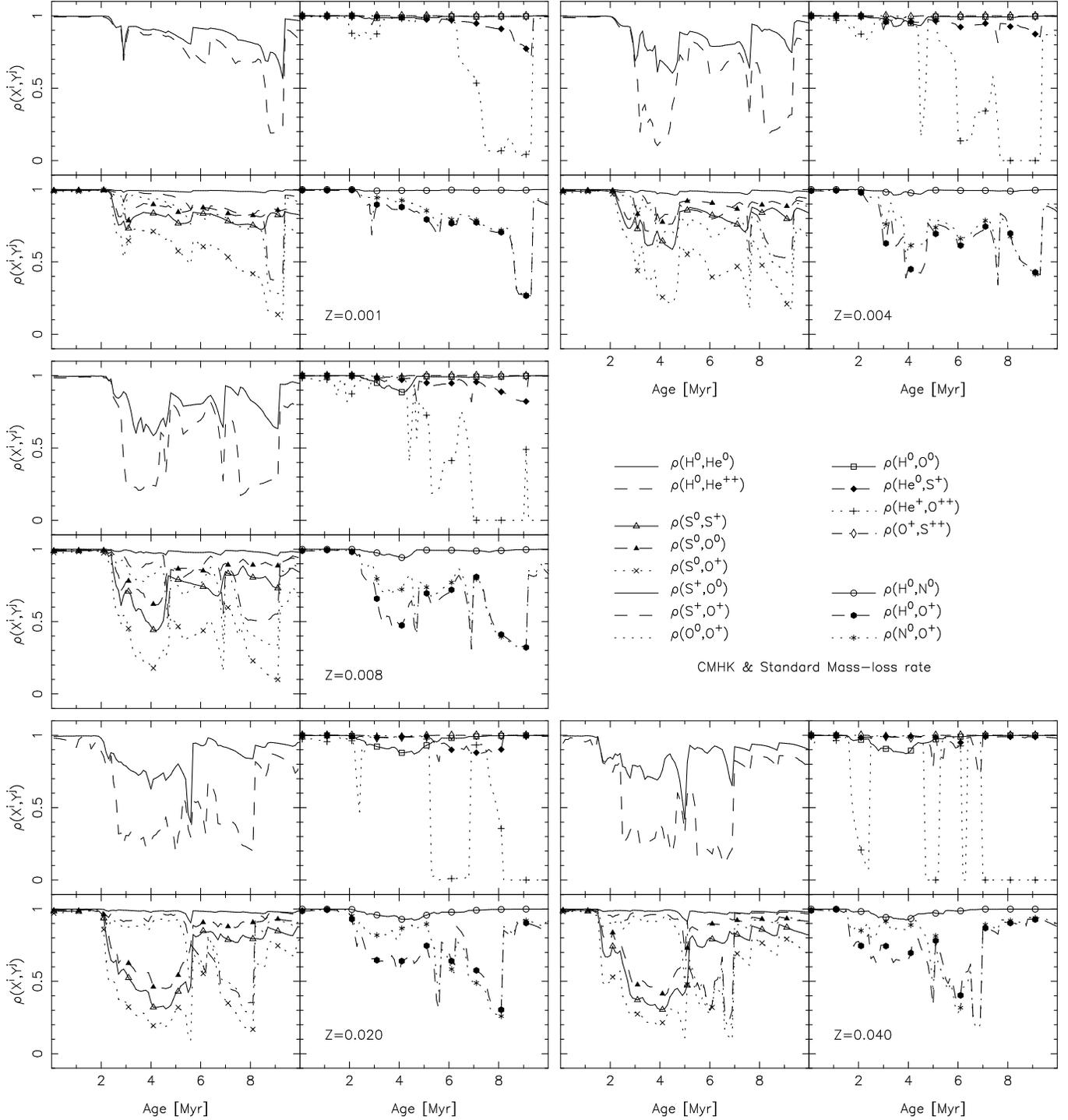}}
  \caption[]{Linear correlation
coefficient $\rho(X^i,Y^j)$  as a function of time for different
  metallicities and ions.
 The five main panels correspond to the
metallicities $Z$=0.001 (top-left), $Z$=0.004 (top-right), $Z$=0.008
(middle), $Z$=0.020 (solar, bottom-left) and $Z$=0.040 (bottom-right).  Each main panel is further subdivided into
four sub-panels: the top-left sub-panel corresponds to
$\rho(\mathrm{H}^{0},\mathrm{He}^{0})$, and
$\rho(\mathrm{H}^{0},\mathrm{He}^{+})$; the top-right sub-panel corresponds
to the correlation coefficients of ions with similar ionising
potential; the
bottom-left sub-panel corresponds to correlation coefficients related to
the $\eta$ parameter;  
and the bottom-right sub-panel 
to correlation coefficients related to
the 
 [\ion{N}{ii}] 6584/H$\alpha$
vs. [\ion{O}{iii}] 5007/H$\beta$ diagram.
}
\label{fig:cov1}
\end{figure*}

In this section, we will discuss the information that can be obtained
through the study of the correlation coefficients 
$\rho[Q_{a_\nu}(X^i),Q_{a_\nu}(Y^j)]$ between
the effective rate of photons of the ions $X^i$ and $Y^j$.
For the sake of simplicity, we will use in the following the notation
 $\rho(X^i,Y^j)\equiv \rho(Q_{a_\nu}(X^i),Q_{a_\nu}(Y^j))$.

As indicated in Section~\ref{sec:sampeffects}, 
a linear correlation coefficient
$\rho(X^i,Y^j)$ close to unity indicates that
$Q_{a_\nu}(X^i)$ and $Q_{a_\nu}(Y^j)$
are essentially produced by
the same stars within a stellar population. 
Note that the correlation between observable quantities 
is not affected by sampling effects in the IMF;
nevertheless, we take advantage of the statistical formalism introduced 
in Section~\ref{sec:sampeffects} for the analysis of the effects
of incomplete sampling to compute the values of the correlation coefficients.

The correlation coefficients between effective rates of ionising photons
may provide useful insights into
the concept of ionisation correction factors ({\sl icf}s) 
used in abundance determinations.
A correlation coefficient
close to one is a necessary condition to relate the 
abundance of two different ions in a ionised nebula
independently of the ionising stellar population.
The assumption of the existence of general recipes 
relating the abundances of two ions
is conceptually analogous to the one underlying
the definition of the {\sl icf}s. 
However, a correlation coefficient close to one 
is not a sufficient condition for a correct 
definition of the {\sl icf}s, since the
properties of the gas might also be relevant
in the determination of the ionisation structure of a nebula
\citep{L03}.

The correlation coefficients between the effective rates of ionising photons
of different ions may also be used
to make predictions on diagnostic diagrams.
However,
in the case of diagnostic diagrams an additional difficulty appears,
because they do not relate absolute intensities but ratios. Hence, e.g.,
the correlation coefficients
$\rho(\mathrm{H}^0,\mathrm{N}^{0})$ and
$\rho(\mathrm{H}^0,\mathrm{O}^{+})$ are not sufficient 
to derive a correlation in the  [\ion{N}{ii}] 6584/H$\alpha$
vs. [\ion{O}{iii}] 5007/H$\beta$ diagram, 
since it is also necessary to know
$\rho(\mathrm{N}^0,\mathrm{O}^{+})$. 
In addition, even in the case of a full
linear correlation ($\rho=1$), 
the tilt of the regression line and the position
of the mean value are required, so a photoionisation model is always
needed.  For the time being, we will only obtain the correlation
coefficients, to see whether correlations exist. In Paper III 
we will apply these results to make predictions on 
the dispersion in the diagnostic diagram.

The results of the calculation of the correlation coefficients are shown in
Fig.~\ref{fig:cov1} for selected ions and diagnostic diagrams.  It is
interesting to compare this figure with Fig.~\ref{fig:Mminsed}, where
${\cal{M}}^{min}[Q_{a_\nu}(X^{i})]$ is given as a function of time for
different metallicities and ions. A similar behaviour is observed in both
figures: the larger the difference of the
${\cal{M}}^{min}[Q_{a_\nu}(X^{i})]$ values, the lower the correlation
coefficient. This is easily understood: a larger difference in
${\cal{M}}^{min}$ means that the stars that contribute the most to the
corresponding effective rates of ionising photons are different, and hence,
the effective rates are loosely correlated.

In general, all correlation coefficients are close to unity until the WR
phase begins. This phase introduces extra ionising photons associated with
a handful of luminous stars, and hence destroy the correlations
between the effective rates of ionising photons.  
Correlation coefficients closer to one are preferentially found
at low metallicities, due to the paucity of WR stars.

Fig. \ref{fig:cov1} is divided into five main panels corresponding to
different metallicities.  Each main panel is further subdivided into
four sub-panels: the top-left sub-panel shows
$\rho(\mathrm{H}^{0},\mathrm{He}^{0})$, and
$\rho(\mathrm{H}^{0},\mathrm{He}^{+})$, the top-right sub-panel shows
the correlation coefficients of ions with similar ionising
potential, the
bottom-left sub-panel shows the correlation coefficients related to
the $\eta$ parameter,
and the bottom-right sub-panel 
the correlation coefficients related to
the 
 [\ion{N}{ii}] 6584/H$\alpha$
vs. [\ion{O}{iii}] 5007/H$\beta$ diagram.

The $\rho(\mathrm{H}^{0},\mathrm{He}^{0})$ and
$\rho(\mathrm{H}^{0},\mathrm{He}^{+})$ correlation coefficients have values
significatively different from 1 starting from the onset of the WR-phase. The larger
the metallicity, the lower the correlation coefficient, implying that
the ratios $Q(\mathrm{He}^0)/Q(\mathrm{H}^0)$ and
$Q(\mathrm{He}^+)/Q(\mathrm{H}^0)$ are not robust constraints of the general
properties of a cluster (like the upper mass limit in the IMF, the IMF
slope, and the age): they are quite dependent on the particular stellar
populations in the cluster and their value can strongly vary from cluster
to cluster, especially for poorly populated high metallicity regions
\cite[see, for example,][]{BK02}. On the
other hand, for the same
reason they can be used to infer the relative abundance of
particular populations in a given observed cluster.

The correlation coefficients shown in the top-right sub-panels are
$\rho(\mathrm{H}^{0},\mathrm{O}^{0})$ (the two ions having 
ionising potentials of 13.599 and 13.618 eV respectively), 
$\rho(\mathrm{He}^{0},\mathrm{S}^{+})$ (ionising potentials 
of 24.588 and 23.330 eV respectively),
$\rho(\mathrm{He}^{+},\mathrm{O}^{++})$ (ionising potentials 
of 54.418 and 54.936 eV respectively),
and $\rho(\mathrm{O}^{+},\mathrm{S}^{++})$ (ionising
potentials of 35.118 and 34.830 respectively).   
With the only exception of $\rho(\mathrm{He}^+,\mathrm{O}^{++})$, 
the correlation coefficients are pretty close to 1 in all the cases,
as expected during the WR phase.  
In contrast, the correlation coefficient of
$\rho(\mathrm{He}^+,\mathrm{O}^{++})$ does not have a constant value, but
surprisingly varies between values close to one and values close to zero.
The zero value does not necessarily imply that the two quantities are not
correlated in the corresponding age ranges: it rather indicates that one or
both of the $Q_{a_\nu}$ values are zero, that is
$\rho(\mathrm{He}^+,\mathrm{O}^{++})=0$ implies the absence of photons
above 54.418 eV in the model atmospheres of stars at the corresponding ages.

The bottom-left sub-panels correspond to correlation coefficients related
to the $\eta$ parameter, that is the ratio of $\mathrm{O}^{+}/
\mathrm{O}^{++}$ to $\mathrm{S}^{+}/ \mathrm{S}^{++}$ \citep{VP88}. Note
that the $\eta$ parameter is measured from forbidden lines, thus it is
related to $Q_{a_\nu}(\mathrm{O}^{0})$, $Q_{a_\nu}(\mathrm{O}^{+})$,
$Q_{a_\nu}(\mathrm{S}^{0})$ and $Q_{a_\nu}(\mathrm{S}^{+})$.  In general,
the correlation coefficients involving $\mathrm{S}^{0}$ are markedly lower
than 1. Remember that $\mathrm{S}^{0}$ was the ion with the lowest
${\cal{M}}^{min}$, which also means that it is produced by a population of
stars different from those responsible for the ionising flux (in fact, it
is produced by the ionising stars {\it plus} other stars): this fact
naturally produces a poor correlation.  As in the case of
$\rho(\mathrm{H}^{0},\mathrm{He}^{0})$ and
$\rho(\mathrm{H}^{0},\mathrm{He}^{+})$, the $\eta$ parameter is not a good
parameter to constrain the age of the cluster, but it is a good parameter
to determine the relative proportion of non-ionising and ionising stars,
and the softness of the ionising radiation for specific clusters.

Finally, the interesting correlation coefficients
$\rho(\mathrm{H}^0,\mathrm{O}^{+})$ and
$\rho(\mathrm{N}^0,\mathrm{O}^{+})$, together with
$\rho(\mathrm{H}^0,\mathrm{N}^{0})$, can be used to
estimate the dispersion in the [\ion{N}{ii}] 6584/H$\alpha$
vs. [\ion{O}{iii}] 5007/H$\beta$ diagram. 
The effective rates of ionising photons
$Q_{a_\nu}(\mathrm{H}^0)$ and $Q_{a_\nu}(\mathrm{N}^{0})$ are strongly
correlated,
implying that the abundances of the two ions are linearly dependent.
In turn, this relation suggests that both elements will show 
a similar correlation with O$^{+}$, 
a fact confirmed by 
$\rho(\mathrm{H}^0,\mathrm{O}^{+})$ 
presenting a behaviour similar to 
$\rho(\mathrm{N}^0,\mathrm{O}^{+})$.
This has an additional implication: the position of data points
in the [\ion{N}{ii}] 6584/H$\alpha$
vs. [\ion{O}{iii}] 5007/H$\beta$ diagram depends
on two variables only (either O$^{+}$ and H$^0$ or O$^{+}$ and N$^{0}$), 
and so the data points will lie on a uniparametric line rather
than being spread on the entire plane. 
Hence, the correlation coefficients obtained through the analysis
of the sampling effects may explain the fact that 
observational data follow a narrow sequence. 
However, the sequence is uniparametric only for a given
metallicity, and it could be different when metallicity varies,
yielding a larger scatter in the diagnostic diagram. That is, in fact, the
situation in the lower region of this diagram, which 
is populated by data points corresponding to different metallicities.

\section{Conclusions}\label{sec:conclusions}

In this paper we have presented an evaluation of the dispersion 
in the properties of the ionising flux obtained from synthesis models 
due to the IMF sampling.
The ionising fluxes have been characterised by 
appropriate physical quantities, the effective rates of ionising photons,
which are obtained from the integration of
the ionising flux weighed by the photoionisation cross section 
of different ions, with the future goal of using them in the
evaluation of sampling effects on emission line spectra.

We have obtained the effective rate of ionising photons for a wide set of
metallicities and atmosphere models, comparing
the results of two different codes to assess some of the
systematic effects. The analysis
of the effective rates 
obtained with different model atmospheres is consistent with
previous studies on the impact of the atmosphere models on the ionising
flux of star forming regions, and it additionally 
provides a proper statistical ground for such assessments. 
We have identified  problems in the
computation of the ionising flux in regimes 
where  stellar evolution does not present a continuous behaviour
with the initial stellar mass, such as at the end of the WR phase. 

We have computed the corresponding minimum initial cluster masses
for which the ionising continuum obtained by synthesis models can be
applied. The minimum masses range from 10$^3$ to more than 10$^6$
M$_\odot$ depending on the metallicity and the age of the stellar population.
For observed clusters with stellar masses below the minimum mass, the
ionising flux is better accounted for by a single star,
rather than by a population of massive stars, so that 
the use of synthesis models is not appropriate in these cases.
We have also evaluated the dispersion in the
distribution of the effective rates of ionising photons at 
the minimum mass, and found it to be around 60\%. 
This means that the initial mass of 
observed stellar clusters must be larger than
a minimum value in the range $3\times 10^{4}$ M$_\odot$ to 
$3\times 10^{7}$ M$_\odot$, depending on 
age and metallicity,
in order for evolutionary
synthesis models that assume a completely sampled IMF 
to be applied.
We have also found indications suggesting 
that the emission of [\ion{S}{ii}]
is much less affected by sampling effects than that of other ions. The 
\ion{He}{ii} lines are in turn the most affected by sampling, 
especially during the WR phase, and should not be used to constrain the
evolutionary status of stellar clusters.

We also studied the correlations between different effective rates of
ionising photons, finding in some cases correlation coefficients close to
one.  This result agrees with the observational finding that \ion{H}{ii}
regions over a vast range of scales, ranging from those ionised by single
stars to those ionised by super-stellar clusters, are found in a relatively
narrow band in some emission-line diagnostic diagrams.

Tables containing the effective rate of ionising photons, their
corresponding $\cal{N}$, the corresponding LLL, and the complete
correlation matrix for H$^{0}$, He$^{0}$, He$^{+}$, N$^{0}$, N$^{+}$,
N$^{++}$, O$^{0}$, O$^{+}$, O$^{++}$, S$^{0}$, S$^{+}$, S$^{++}$, C$^{0}$,
Ne$^{0}$, Ne$^{+}$, Ar$^{0}$, Ar$^{+}$, and Ar$^{++}$ ions are available at
the {\sc www} address {\tt http://laeff.inta.es/users/mcs/SED/}.  They
include the results using the CMHK code with standard and high mass-loss
rate\footnote{The resulting values in those tables are normalised to 1
M$_\odot$ for a Salpeter IMF in the mass range 2--120 M$_\odot$ instead the
0.1--120 M$_\odot$ mass range used in this paper.}.  

\begin{acknowledgements}
We want to acknowledge the referee, L. Smith, for very useful
comments. We want also acknowledge G. Ferland for his clear and extensive
documentation of {\tt Cloudy}.
MC has been partially supported by the Project AYA3939-C03-01 of the
Spanish Programa Nacional de Astronom\'\i a y Astrof\'\i sica of the MCyT.  
VL is
supported by a Marie Curie Fellowship of the European Community programme
{\it Improving Human Research Potential and the Socio-Economic Knowledge
Base} under contract number HPMF-CT-2000-00949.
\end{acknowledgements}

\bibliographystyle{apj}
\end{document}